\documentclass{aastex}
\usepackage{spr-astr-addons}
\usepackage{url}

\RequirePackage{color}

\begin{document}

\title{Geodesic flows around charged black holes in two dimensions}
\slugcomment{Not to appear in Nonlearned J., 45.}
\shorttitle{Short article title}
\shortauthors{Autors et al.}

\author{RAVI SHANKAR KUNIYAL\altaffilmark{1}} \and \author{RASHMI UNIYAL\altaffilmark{2}}
\and \author{HEMWATI NANDAN\altaffilmark{3}}
\affil{Department of Physics, Gurukul Kangri Vishwavidyalaya,\\
Haridwar, Uttarakhand 249 407, India}
\email{ravikuniyal09@gmail.com}
\and
\author{ALIA ZAIDI\altaffilmark{4}}
\affil{School of Physics and Center for Theoretical Physics,
University of Witwatersrand, Johannesburg, Wits 2050, South Africa}

\altaffiltext{1}{ravikuniyal09@gmail.com}
\altaffiltext{2}{rashmiuniyal001@gmail.com}
\altaffiltext{3}{hnandan@iucaa.ernet.in}
\altaffiltext{4}{Alia.Zaidi@wits.ac.za}

\begin{abstract}
We study the evolution of timelike geodesics for two dimensional black hole spacetimes arising in string theory and general theory of relativity by solving the Raychaudhuri equation  for expansion scalar as an \textit{initial value problem}.
The possibility of geodesic focusing/defocusing is then examined accordingly with different settings of black hole parameters.
In view of the geodesic focusing/defocusing, the critical value of expansion scalar is also calculated in each case.
The effect of the charge and the cosmological constant on the evolution of expansion scalar for timelike geodesics is discussed in detail.
\end{abstract}



\section{Introduction}

The classical black hole (BH) solutions in general relativity (GR) in four dimensions ($4$D) are endowed with new features, if the theory is modified by low energy string corrections in the action (\citeauthor{gar91} \citeyear{gar91}; \citeauthor{cam90} \citeyear{cam90}; \citeauthor{shp91} \citeyear{shp91} ). Action for a two dimensional ($2$D) gravity can be obtained if one introduces a scalar field (whose logarithm is called dilaton in the notions of string theory).
Non-minimal coupling of the dilaton and axion fields with gravity and other fields is one of the key aspects of these modified actions(\citeauthor{bek72} \citeyear{bek72}; \citeauthor{cha70} \citeyear{cha70}).
The dynamics of gravity for a generic higher dimensional theory may therefore be treated as an effective theory at lower dimensions by the introduction of dilaton fields.\\
As it is well known that if mass of the BH is large compared to the Planck mass then all vacuum solutions of Einstein gravity are basically the approximate solutions of low-energy string gravity.
The charged BH solutions emerging in string theory in which the dilaton field couples to the Yang-Mills field are fundamentally different from the classical Reissner-Nordstr$\ddot{o}$m BHs (\citeauthor{mig93} \citeyear{mig93}) in GR.\\
To understand the gravitational field around the BH (\citeauthor{poi04} \citeyear{poi04}; \citeauthor{wal84} \citeyear{wal84}; \citeauthor{har03} \citeyear{har03}; \citeauthor{sch85} \citeyear{sch85}; \citeauthor{cha83} \citeyear{cha83}; \citeauthor{jos97} \citeyear{jos97}), the initial step may be to study the geodesic motion for given test particles in a given spacetime.
There are various interesting aspects related to the geodesic motion around a given $2$D BH spacetime, including the energy distribution (\citeauthor{vag03} \citeyear{vag03}), the varying constants problem(\citeauthor{vag3} \citeyear{vag3}), the kinematics of timelike as well as null geodesic congruences(\citeauthor{das09} \citeyear{das09}; \citeauthor{das08} \citeyear{das08}; \citeauthor{das9} \citeyear{das9}; \citeauthor{gho10} \citeyear{gho10}; \citeauthor{fer12} \citeyear{fer12}; \citeauthor{pug11} \citeyear{pug11}) and geodesic deviation(\citeauthor{kol03} \citeyear{kol03}; \citeauthor{uni14} \citeyear{uni14}; \citeauthor{runi14} \citeyear{runi14}).
Further the evolution of the kinematical quantities (i.e. isotropic expansion, shear and rotation) along the geodesic flows, are governed by the Raychaudhuri equations (\citeauthor{kar08} \citeyear{kar08}; \citeauthor{kar07} \citeyear{kar07}; \citeauthor{dad05} \citeyear{dad05}; \citeauthor{ehl07} \citeyear{ehl07}).
These equations result from relating the evolution of deviation between two neighbouring geodesics to the curvature of spacetime.\\
The structure of the paper is as follows.
A brief introduction to the spacetimes considered is presented in the next section.
In section III, the kinematics of deformation is discussed, while the exact solutions for geodesics and geodesic deviation in this $2$D stringy BH background have already been studied in our previous work (\citeauthor{uni14} \citeyear{uni14}).
In $2$D case, the Raychaudhuri equation is just the evolution equation for expansion scalar while the shear and rotation vanishes manifestly.
The evolution equations are analytically solved and critical values for the expansion scalar is calculated for geodesic focusing/defocusing in each case.
In section IV, the role of initial conditions on geodesic focusing/defocusing is analysed and compared with another charged $2$D BH spacetime arising in GR.
The results are summarized in section V along with the conclusions.

\section{The Model and Spacetimes}
The general spacetime for spherically reduced $2$D BHs is given as,
\begin{equation}
{ds}^2=-X(r)\hspace{0.1cm}dt^2+ X(r)^{-1}\hspace{0.1cm}dr^2 ,
\label{eqn:gen_spacetime}
\end{equation}
for specific choice of X(r) emerging in GR and other alternative theories of gravity viz. string theory. To understand the dynamics of a test particle in the spacetime given in eq.(\ref{eqn:gen_spacetime}), the general structure of the geodesic equations corresponding to the above line element is given by,
\begin{eqnarray}
\ddot{t} + X^{'}(r)\hspace{0.1cm}X(r)^{-1}\hspace{0.1cm}\dot{r}\hspace{0.1cm}\dot{t}=0,\nonumber\\
\label{eqn:geod_eqn for t}
\ddot{r}+\frac{1}{2}\hspace{0.1cm}X(r)\hspace{0.1cm}X^{'}(r)\hspace{0.1cm}\dot{t}^2-\frac{1}{2}\hspace{0.1cm}X^{'}(r)X^{-1}(r)\hspace{0.1cm}\dot{r}^2=0,
\label{eqn:geod_eqn for r}
\end{eqnarray}
where prime ($\prime$) and dot ($\cdot$) represent the differentiation with respect to $r$ and affine parameter ($\tau$) respectively.
The general form of the first integral of the geodesic equations (using the constraint equation as $g_{\alpha\beta}u^{\alpha}u^{\beta}=-1$ for timelike geodesics) for above case can be obtained as,
\begin{flushleft}
\begin{eqnarray}
\dot t = E\hspace{0.1cm}X^{-1}(r),\nonumber\\
\label{eqn:first_int_01}
\vspace{1cm}
{\dot r} = \pm\sqrt{E^2-X(r)},
\label{eqn:first_int_02}
\end{eqnarray}
\end{flushleft}
 where $E$ is a constant. We are considering the following BH spacetimes for our investigation,\vspace{5mm}\\
 \textbf{Case I}: The line element corresponding to the $2$D charged stringy BH, arising from heterotic string theory (\citeauthor{mcg92} \citeyear{mcg92}) in the ``Schwarzschild'' gauge is given as,\\
\begin{equation}
X(r)=1-2\hspace{0.1cm}m\hspace{0.1cm}e^{-2\lambda r}+{q_{_S}^2}\hspace{0.1cm}e^{-4\lambda r},
\label{eqn:stringy_spacetime}
\end{equation}
\noindent where, $0 < t < +\infty$, $r_+ < r < +\infty$, $r_+$ is the event horizon of the BH.
The parameters $m$, $q_{_S}$ and $\lambda$ are the mass, electric charge and cosmological constants of the BH respectively and horizons are placed at,
\begin{equation}
r_{\pm}=\frac{1}{2\lambda}\ln{\left[\frac{q_{_S}^2}{m\pm\sqrt{m^2-{q_{_S}^2}}}\right]} .
\label{stringy_horizon}
\end{equation}
 The asymptotic flat region is $r=+\infty$ (since $q_{_S}$ and $\lambda$ are positive)
 and the curvature singularity is at $r=-\infty$.\\
First integrals of geodesic equations given in eq.(\ref{eqn:first_int_02}) reduce to the following form for the above spacetime specified by the eq.(\ref{eqn:stringy_spacetime}),
\begin{eqnarray}
\dot{t}=\frac{E}{1- 2\hspace{0.1cm}m\hspace{0.1cm}e^{-2\lambda r}+q^2_{_S}\hspace{0.1cm}e^{-4\lambda r}},
\nonumber \\
\label{eqn:first_int_t for stringy}
\dot{r}= \sqrt{E^2-1 + 2\hspace{0.1cm}me^{-2\lambda r}- q^2_{_S}\hspace{0.1cm}e^{-4\lambda r}}.
\label{eqn:first_int_r for stringy}
\end{eqnarray}
\vspace{5mm}\\
\textbf{Case II}: The metric (\ref{eqn:RN_spacetime}) represents usual charged BH (\citeauthor{mcg92} \citeyear{mcg92}) in $2$D, i.e. the two dimensional analogue of $4$D Reissner-Nordstr$\ddot{o}$m BH spacetime with,
\begin{equation}
X(r)=1-\frac{2\hspace{0.1cm}m}{r}+\frac{q_{_{RN}}^2}{r^2},
\label{eqn:RN_spacetime}
\end{equation}
 where the asymptotic flat region is at $r=\infty$, and the curvature singularity is at $r=0$.
 Here $m$ and $q_{_{RN}}$ are the mass and electric charge of the BH. The positions of horizons are given as,
\begin{equation}
 r_\pm=\frac{q_{_{RN}}^2}{m\pm\sqrt{m^2-q_{_{RN}}^2}} .
 \label{RN_horizon}
\end{equation}
First integrals of the geodesic equations (\ref{eqn:first_int_02}), now reduce to the following form for spacetime given in eq.(\ref{eqn:RN_spacetime}),
\begin{eqnarray}
\dot{t}=\frac{E}{1-\frac{2\hspace{0.1cm}m}{r}+\frac{q^2_{_{RN}}}{r^2}} ,
\nonumber\\
\label{eqn:first_int_rn_01}
\vspace{1cm}
\dot{r}= \pm\sqrt{E^2-(1-\frac{2\hspace{0.1cm}m}{r}+\frac{q^2_{_{RN}}}{r^2})} .
\label{eqn:first_int_rn_02}
\end{eqnarray}
There are two singularities present for both of the spacetimes given by eq.(\ref{eqn:stringy_spacetime}) and eq.(\ref{eqn:RN_spacetime}).
For extremal case in which charge parameter is equal to the mass, both singularities are merged, hence only one event horizon exists.
It is worth noticing that the case where charge dominates over mass represents a naked singularity (i.e. a singularity without event horizon).\\
\section{Kinematics of Deformation}
Now the first approach to have an expression for the expansion scalar ($\theta$) of timelike geodesics is to calculate $\theta$ as a function of $r$ from the expression of the corresponding first integral equation $u^i=(\dot{t}, \dot{r})$ and using $\theta=\nabla_iu^i$.
The expression of $\theta$ for spacetime given by eq.(\ref{eqn:stringy_spacetime}) using eqs.(\ref{eqn:first_int_r for stringy}) reads,
\begin{equation}
\theta_{_S}=-\frac{X^{'}(r)}{2\hspace{0.1cm}\dot{r}}=\frac{-2\hspace{0.1cm}m\hspace{0.1cm}\lambda\hspace{0.1cm}e^{-2 \lambda r}+2\hspace{0.1cm}q^2_{_S}\hspace{0.1cm}e^{-4 \lambda r}}{\pm\hspace{0.1cm}\sqrt{E^2 - 1 + 2\hspace{0.1cm}m\hspace{0.1cm}e^{-2 \lambda\hspace{0.1cm}r}-q^2_{_S}\hspace{0.1cm}e^{-4 \lambda\hspace{0.1cm}r}}}.\\
\label{eqn:theta_stringy}
\end{equation}
This expression for $\theta$ diverges to infinity at a turning point of the geodesic motion (i.e. $\dot{r}=0$).
But  whether geodesics will focus or defocus depends on the implicit relation between different parameters given in numerator of the expression as well as on the sign of radial velocity, $\dot{r}$.
For $\dot{r}\rightarrow 0^+$ focusing of geodesics will occur if $m\hspace{0.1cm}\lambda > q^2_{_S}\hspace{0.1cm}e^{-2\lambda\hspace{0.1cm}r}$, while geodesics will defocus if $m\hspace{0.1cm}\lambda\hspace{0.1cm}e^{2\lambda r} < q^2_{_S}$.
Similarly, for $\dot{r}\rightarrow 0^-$ one will have geodesic focusing with $m\hspace{0.1cm}\lambda\hspace{0.1cm} e^{2\lambda r} < q^2_{_S}$ and defocusing with $m\hspace{0.1cm}\lambda > q^2_{_S}\hspace{0.1cm}e^{-2\lambda r}$.\\
While the expression of $\theta$ for spacetime given by eq.(\ref{eqn:RN_spacetime}) using eqs.(\ref{eqn:first_int_rn_02}) is given as,
\begin{equation}
\theta_{_{RN}}=-\frac{X^{'}(r)}{2\hspace{0.1cm}\dot{r}}=\frac{-m\hspace{0.1cm}r + q^2_{_{RN}}}{\pm\hspace{0.1cm} r^2\hspace{0.1cm}\sqrt{E^2\hspace{0.1cm}r^2 - r^2 + 2\hspace{0.1cm}m\hspace{0.1cm}r - q^2_{_{RN}}}}.
\label{eqn:theta_rn}
\end{equation}
This expression for $\theta$ also diverges to infinity at $\dot{r}=0$.
Here again condition for geodesic focusing or defocusing is given by the implicit relation between different parameters given in numerator of the expression as well as on the sign of $\dot{r}$.
For $\dot{r}\rightarrow 0^+$ focusing of geodesics will occur if $m\hspace{0.1cm}r > {q^2_{_{RN}}}$, while geodesics will defocus if $m\hspace{0.1cm}r < {q^2_{_{RN}}}$.
Similarly, for $\dot{r}\rightarrow 0^-$ one will have geodesic focusing with $m\hspace{0.1cm}r<{q^2_{_{RN}}}$ and defocusing with $m\hspace{0.1cm}r > {q^2_{_{RN}}}$.
An explicit dependence of $\theta$ on $\tau$ can be obtained by substituting $r(\tau)$ from the solution of corresponding geodesic equations for $r$ given in eq.(\ref{eqn:first_int_r for stringy}) and eq.(\ref{eqn:first_int_rn_02}).
In the expressions of $\theta$ given in eq. (\ref{eqn:theta_stringy}) and eq. (\ref{eqn:theta_rn}), there is no way to specify initial conditions on $\theta$. Hence the effect of initial conditions on the evolution of geodesic congruences cannot be discussed using these expressions.
So in the next step to analyse the behaviour of geodesics congruences on imposing the initial conditions we will consider the Raychaudhuri equation for expansion scalar along with the geodesic equation derived for the above mentioned spacetimes.
\subsection{The Raychaudhuri equation for expansion scalar}
The evolution equations for timelike geodesics for $2$D BH spacetimes consist of the Raychaudhuri equation for expansion scalar (as shear and rotation both vanish for $2$D metrices (\citeauthor{das09} \citeyear{das09})) and the geodesic equation derived for a particular metric.
The Raychaudhuri equation for expansion scalar for general $2$D BH spacetime (\citeauthor{das09} \citeyear{das09}) given in eq.(\ref{eqn:gen_spacetime}) is given by,
\begin{equation}
\dot{\theta}+\theta^2=-R^i_{lim}u^lu^m=-R_{lm}u^lu^m.
\label{eqn:gen_Raychaudhuri_00}
\end{equation}
The general form of eq.(\ref{eqn:gen_Raychaudhuri_00}) can also be written as,
\begin{equation}
\dot{\theta}+{\theta^2}-\frac{R}{2}=0
\label{eqn:gen_Raychaudhuri_01}
\end{equation}
where $R$ is the Ricci Scalar (see \textbf{Appendix I}).
The general Raychaudhuri equation for expansion scalar given in eq.(\ref{eqn:gen_Raychaudhuri_01}) reduces to the following forms for the spacetimes used for the present study:\\
the general Raychudhuri equation given in eq.(\ref{eqn:gen_Raychaudhuri_01}) for the expansion scalar in the background of spacetime given by eq.(\ref{eqn:stringy_spacetime}) for ${E^2}=1$ reads as,
\begin{equation}
\dot{\theta}+\theta^2-\frac{8\hspace{0.1cm}m^2\hspace{0.1cm}\lambda^2}{4\hspace{0.1cm}\tau^2\hspace{0.1cm}\lambda^2 \hspace{0.1cm}m^2 + q^2_{_S}} + \frac{32\hspace{0.1cm}\lambda^2\hspace{0.1cm}m^2\hspace{0.1cm}q^2_{_S}}{(4\hspace{0.1cm}\tau^2\hspace{0.1cm}\lambda^2 \hspace{0.1cm}m^2 + q^2_{_S})^2}=0
\label{eqn:theta_eqn for stringy}
\end{equation}
this equation will now be solved analytically (in next section) for the expansion scalar in view of different charge to mass ratios as, $q_{_S}>>m$, $q_{_S}=m$ and $q_{_S}<<m$ (where $m$ and $q_{_S}$ correspond to mass and charge of the stringy BH).\\
The general Raychudhuri equation given in eq.(\ref{eqn:gen_Raychaudhuri_01}) for the expansion scalar in the background of spacetime given by eq.(\ref{eqn:RN_spacetime}) for ${E^2}=1$ reads as,
\begin{equation}
\dot{\theta}+\theta^2-\frac{16\hspace{0.1cm} m^4}{(q^2_{_{RN}} + (p-\frac{q^2_{_{RN}}}{p})^2)^3} + \frac{48\hspace{0.1cm}{q^2_{_{RN}}}\hspace{0.1cm}m^4}{(q^2_{_{RN}} + (p-\frac{q^2_{_{RN}}}{p})^2)^4}=0
\label{eqn:theta_eqn for rn}
\end{equation}
this equation will now be solved numerically (in next section) for the expansion scalar with conditions $q_{_{RN}}>>m$, $q_{_{RN}}=m$ and $q_{_{RN}}<<m$ (where $m$ and $q_{_{RN}}$ correspond to mass and charge of the Reissner-Nordstr$\ddot{o}$m BH) and $p=(3\hspace{0.1cm}\tau + \sqrt{q_{_{RN}}^6 + 9\hspace{0.1cm}\tau^2\hspace{0.1cm}m^4})^{1/3}$.

\subsubsection{Evolution of expansion scalar for different cases}
In the present section corresponding Raychaudhuri equation (\ref{eqn:theta_eqn for stringy}) for expansion scalar $\theta$ is solved analytically, as an initial value problem with different conditions on BH parameters.
\begin{enumerate}
\item[\textbf{(a)}]\hspace{1mm}\textbf{Charge dominance over Mass i.e. Naked Singularity ($q_{_S}>>m$)};\\
The solution of eq.(\ref{eqn:theta_eqn for stringy}) is given by,
\begin{equation}
\theta(\tau)=\frac{\alpha\hspace{0.1cm}C_1 - 32\hspace{0.1cm} \lambda^2 \hspace{0.1cm}m^2 \hspace{0.1cm} q^2_{_S}\hspace{0.1cm}\tau}{(\beta \hspace{0.1cm} C_1 - 12\hspace{0.1cm}\lambda^2\hspace{0.1cm}m^2\hspace{0.1cm}\tau^2 + q^2_{_S})(4\hspace{0.1cm}\lambda^2\hspace{0.1cm}m^2\hspace{0.1cm}\tau^2 + q^2_{_S})}
\label{eqn:sol_for stringy_qqqm}
\end{equation}
where $\alpha = 16\hspace{0.1cm}\lambda^4\hspace{0.1cm}m^4\hspace{0.1cm}\tau^4 + 24\hspace{0.1cm}\lambda^2\hspace{0.1cm}m^2\hspace{0.1cm}q^2_{_S}\hspace{0.1cm}\tau^2 - 3\hspace{0.1cm}q^4_{_S}$,\\
$\beta = 4\hspace{0.1cm}\lambda^2\hspace{0.1cm}m^2\hspace{0.1cm}\tau^3 -3\hspace{0.1cm}q^2_{_S}\hspace{0.1cm}\tau $.\\
and $C_1$ is the integration constant written in terms of the initial conditions as,
\begin{equation}
C_1=\frac{A\hspace{0.1cm}\theta_0 - 32\hspace{0.1cm}m^2\hspace{0.1cm}q^2_{_S}\hspace{0.1cm}\lambda^2\hspace{0.1cm}\tau_0}{B\hspace{0.1cm}\theta_0 - 16\hspace{0.1cm}m^4\hspace{0.1cm}\lambda^4\hspace{0.1cm}\tau_0^4 - 24\hspace{0.1cm}m^2\hspace{0.1cm}q^2_{_S}\hspace{0.1cm}\lambda^2\hspace{0.1cm}\tau_0^2  + 3\hspace{0.1cm}q^4_{_S}}
\label{eqn:c_for stringy_qqqm}
\end{equation}
\begin{eqnarray}
where,
A = 48\hspace{0.1cm}m^4\hspace{0.1cm}\lambda^2\hspace{0.1cm}\tau_0^4 + 8\hspace{0.1cm}m^2\hspace{0.1cm}q^2_{_S}\hspace{0.1cm}\lambda^2\hspace{0.1cm}\tau_0^2 - q^4_{_S},\hspace{0.1cm}
\nonumber
\\
B= 16\hspace{0.1cm}m^4\hspace{0.1cm}\lambda^5\hspace{0.1cm}\tau_0^5 - 8\hspace{0.1cm}m^2\hspace{0.1cm}q^2_{_S}\hspace{0.1cm}\lambda^2\hspace{0.1cm}\tau_0^3 - 3\hspace{0.1cm}q^4_{_S}\hspace{0.1cm}\tau_0.
\nonumber
\end{eqnarray}
\item[\textbf{(b)}]\textbf{Extremal Black Hole Case ($q_{_S}=m$)};\\
The solution of eq.(\ref{eqn:theta_eqn for stringy}) is given by,
\begin{equation}
\theta(\tau)=\frac{128\hspace{0.1cm}C_2\hspace{0.1cm}\tau^5 + 64\hspace{0.1cm}C_2\hspace{0.1cm}\tau^3 - 12\hspace{0.1cm}\tau^2 + 72\hspace{0.1cm}C_2\hspace{0.1cm}\tau + 3}{(16\hspace{0.1cm}C_2\hspace{0.1cm}\tau^4 + 24\hspace{0.1cm}C_2\hspace{0.1cm}\tau^2 + 3\hspace{0.1cm}\tau - 3\hspace{0.1cm}C_2)(4\hspace{0.1cm}\tau^2 + 1)}
\label{eqn:sol_for stringy_qm}
\end{equation}
where $C_2$ is the integration constant given in terms of the initial conditions as,
\begin{equation}
C_2=\frac{3\hspace{0.1cm}(4\hspace{0.1cm}\theta_0\hspace{0.1cm}\lambda^2\hspace{0.1cm}m^2\hspace{0.1cm}\tau_0^3 - 4\hspace{0.1cm}\lambda^2\hspace{0.1cm}m^2\hspace{0.1cm}\tau_0^2 + \theta_0\hspace{0.1cm}q_{_S}^2\hspace{0.1cm}\tau_0 - q_{_S}^2)}{F\hspace{0.1cm}\theta_0 - 128\hspace{0.1cm}\eta^3\hspace{0.1cm}\tau_0^5 - 64\hspace{0.1cm}\eta^2\hspace{0.1cm}q_{_S}^2\hspace{0.1cm}\tau_0^3 -72\hspace{0.1cm}\eta\hspace{0.1cm}q_{_S}^4\hspace{0.1cm}\tau_0}
\end{equation}
where,\\$F = 64\lambda^6m^6{\tau_0^6}+112\lambda^4m^4q_{_S}^2{\tau_0^4}+12\lambda^2m^2q_{_S}^4{\tau_0^2}-3q_{_S}^6$,
$\eta=\lambda^2\hspace{0.1cm}m^2$.
\item[\textbf{(c)}] \textbf{Mass dominance over Charge ($q_{_S}<<m$)};\\
The solution of eq.(\ref{eqn:theta_eqn for stringy}) is now given as,
\begin{equation}
\theta(\tau)=\frac{4(2m \lambda \tau tan^{-1}(\frac{2 \lambda m \tau}{q_{_S}})+2C_3 \lambda m \tau + q_{_S} \lambda m)}{\kappa+\xi}
\label{eqn:solution for stringy_mmq}
\end{equation}
where, $\kappa=(4 \lambda^2 m^2 \tau^2 + q^2_{_S}) tan^{-1}(\frac{2 \tau \lambda m}{q_{_S}})$
and \\$\xi=4C_3 \lambda^2 m^2 \tau^2 + 2\lambda m q_{_S} \tau + C_3 q^2_{_S}$.
$C_3$ is the integration constant given in terms of the initial conditions as,
\begin{equation}
C_3=-\frac{H\theta_0+J\tau_0-4m\hspace{0.1cm}\lambda\hspace{0.1cm} q_{_S}}{4\hspace{0.1cm}\theta_0\hspace{0.1cm}\lambda^2\hspace{0.1cm}m^2\hspace{0.1cm}\tau_0^2-8m^2\hspace{0.1cm}\lambda^2\hspace{0.1cm}\tau_0 + \theta_0\hspace{0.1cm}q_{_S}^2}
\end{equation}
\begin{eqnarray}
where,
H=\left(4\lambda^2\hspace{0.1cm}m^2\hspace{0.1cm}\tau_0^2 +q_{_S}^2\right) tan^{-1}(\frac{2m\lambda \tau_0}{q_{_S}}),\hspace{0.1cm}
\nonumber
\\
J=2\hspace{0.1cm}\theta_0\hspace{0.1cm}\lambda \hspace{0.1cm}m\hspace{0.1cm}q_{_S}-8m^2\hspace{0.1cm}\lambda^2 \hspace{0.1cm} tan^{-1}(\frac{2m\lambda \tau_0}{q_{_S}}).
\nonumber
\end{eqnarray}
\end{enumerate}
where $\tau_0$ and $\theta_0$ are the initial values of affine parameter and expansion scalar respectively.
\section{Analysis of geodesic focusing and defocusing}
\begin{figure}[h]
\includegraphics[width=3.8cm,height=4cm]{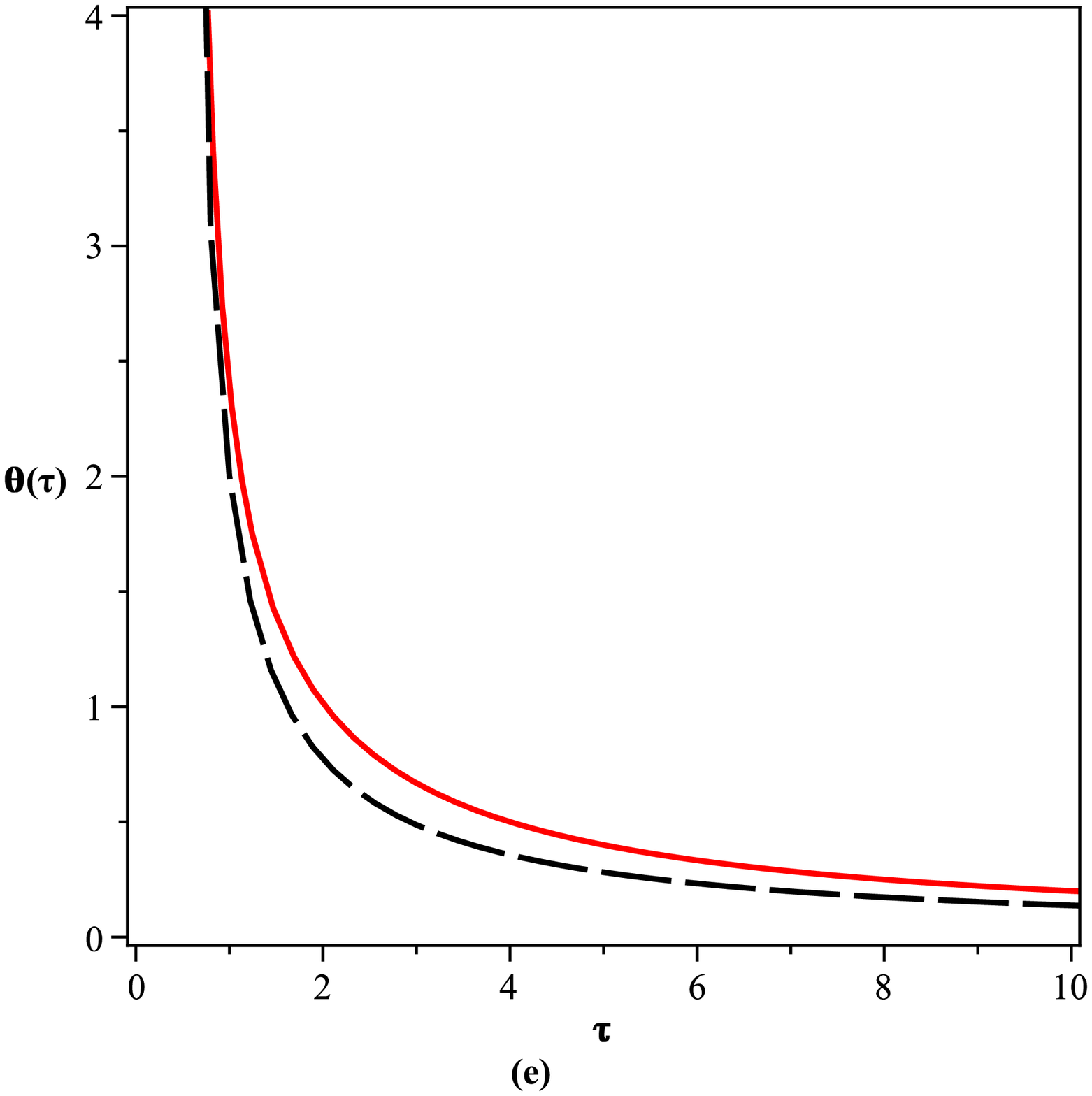}
\hskip3mm\includegraphics[width=3.8cm,height=4cm]{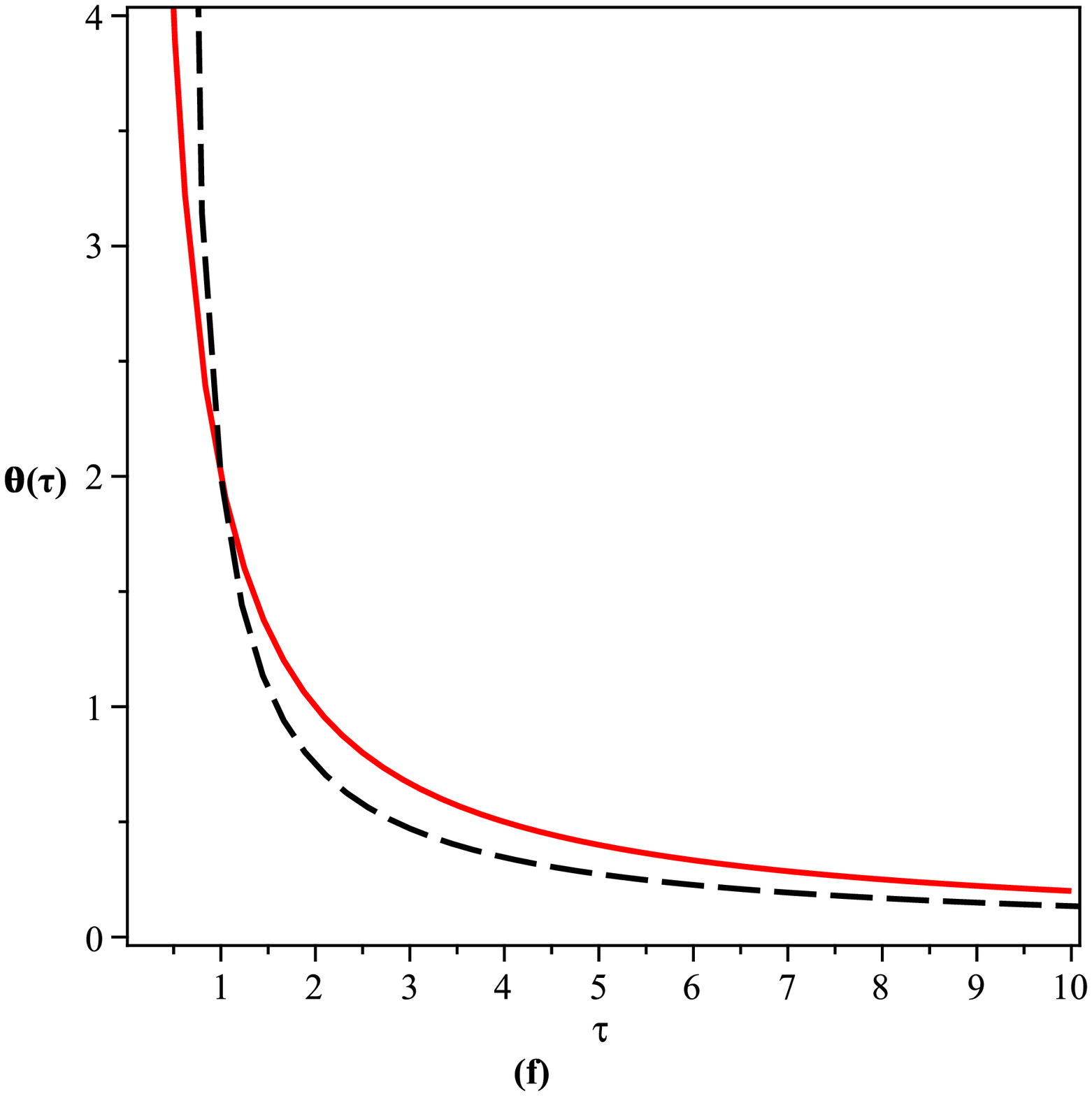}
\vspace{0.5cm}
\caption{(a) $q_{_S}=q_{_{RN}}=100$, $m=1$, $\theta_0^c(300)=-0.02122$; (b) $q_{_S}=q_{_{RN}}=2$, $m=1$ ; (c) $q_{_S}=q_{_{RN}}=m=1$, $\theta_0^c(0.5)=-1.65517$; (d) $q_{_S}=q_{_{RN}}=1$, $m=100$, $\theta_0^c(1)=2$; (e) $q_{_S}=q_{_{RN}}=1$, $m=2$; (f) $q_{_S}=1$, $q_{_{RN}}=0$ (this case corresponds to $2$D Schwarzschild BH), $m=100$; with $\lambda=1$, where solid and dotted curves represent $\theta(\tau)$ for stringy and Reissener Nordstr$\ddot{o}$m BHs respectively.}
\protect\label{f1}
\end{figure}

\begin{figure}[h]
\includegraphics[width=5cm,height=5cm]{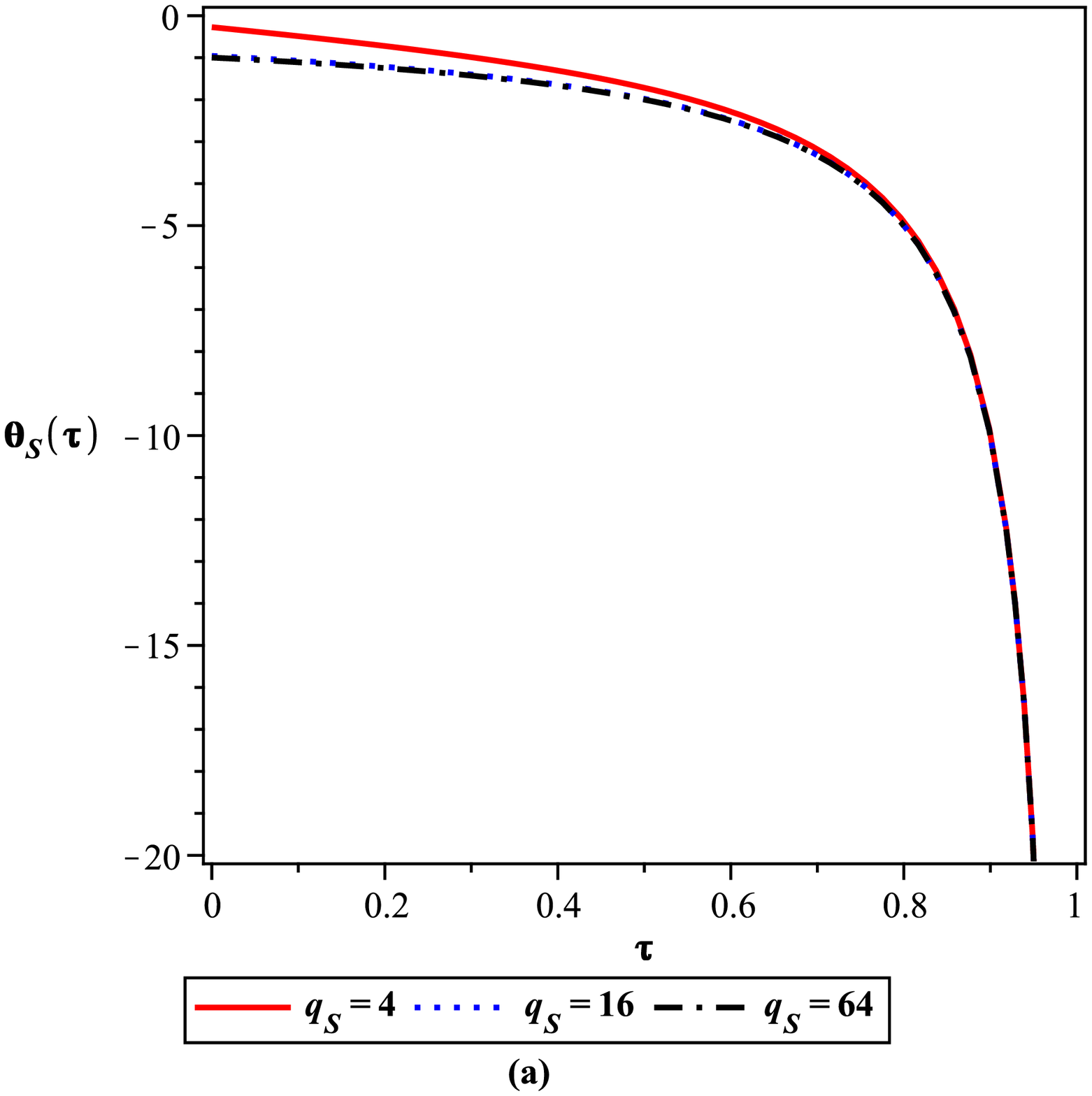} \hskip1mm\includegraphics[width=3cm,height=3cm]{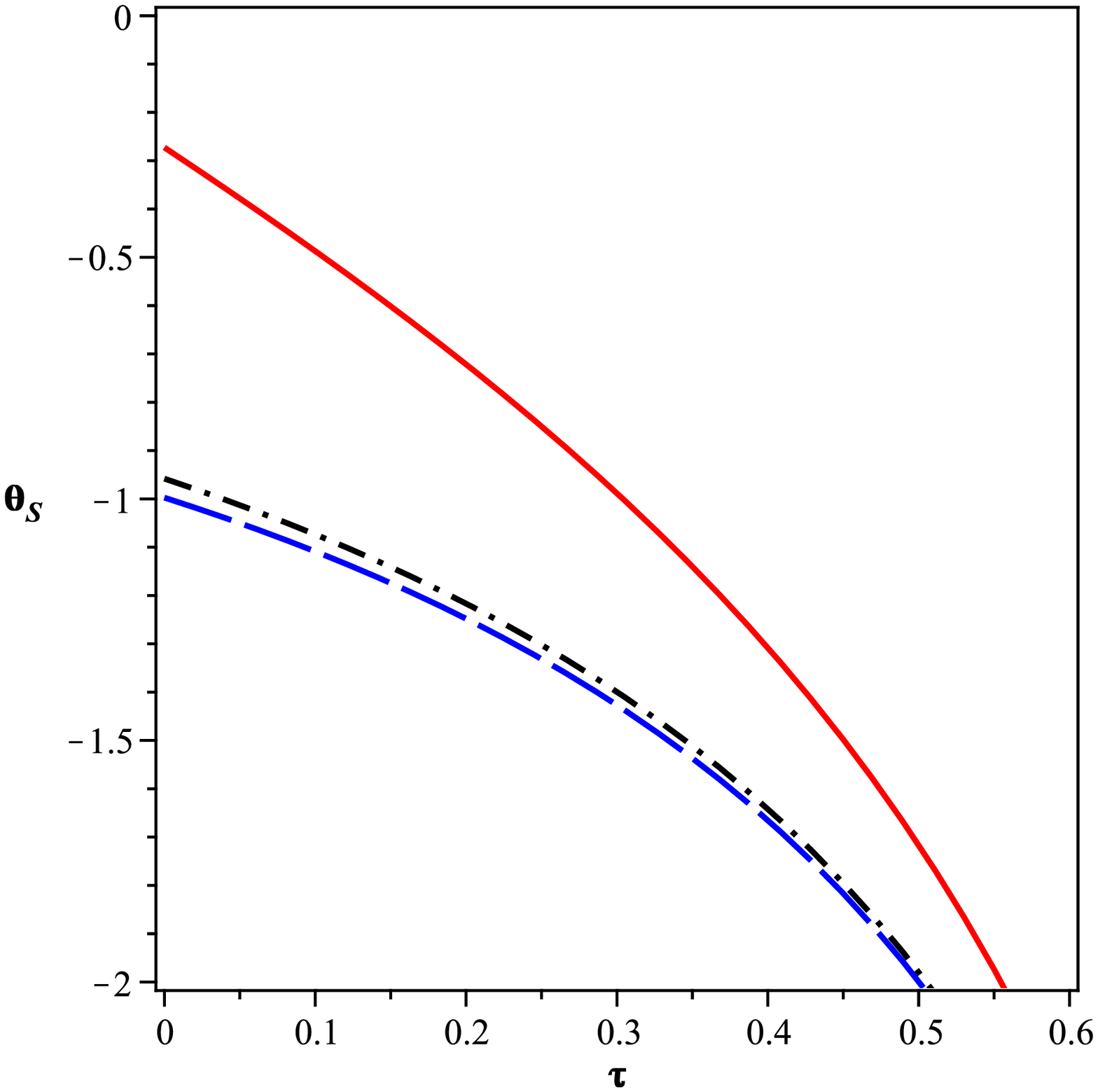}
\hskip1mm \includegraphics[width=5cm,height=5cm]{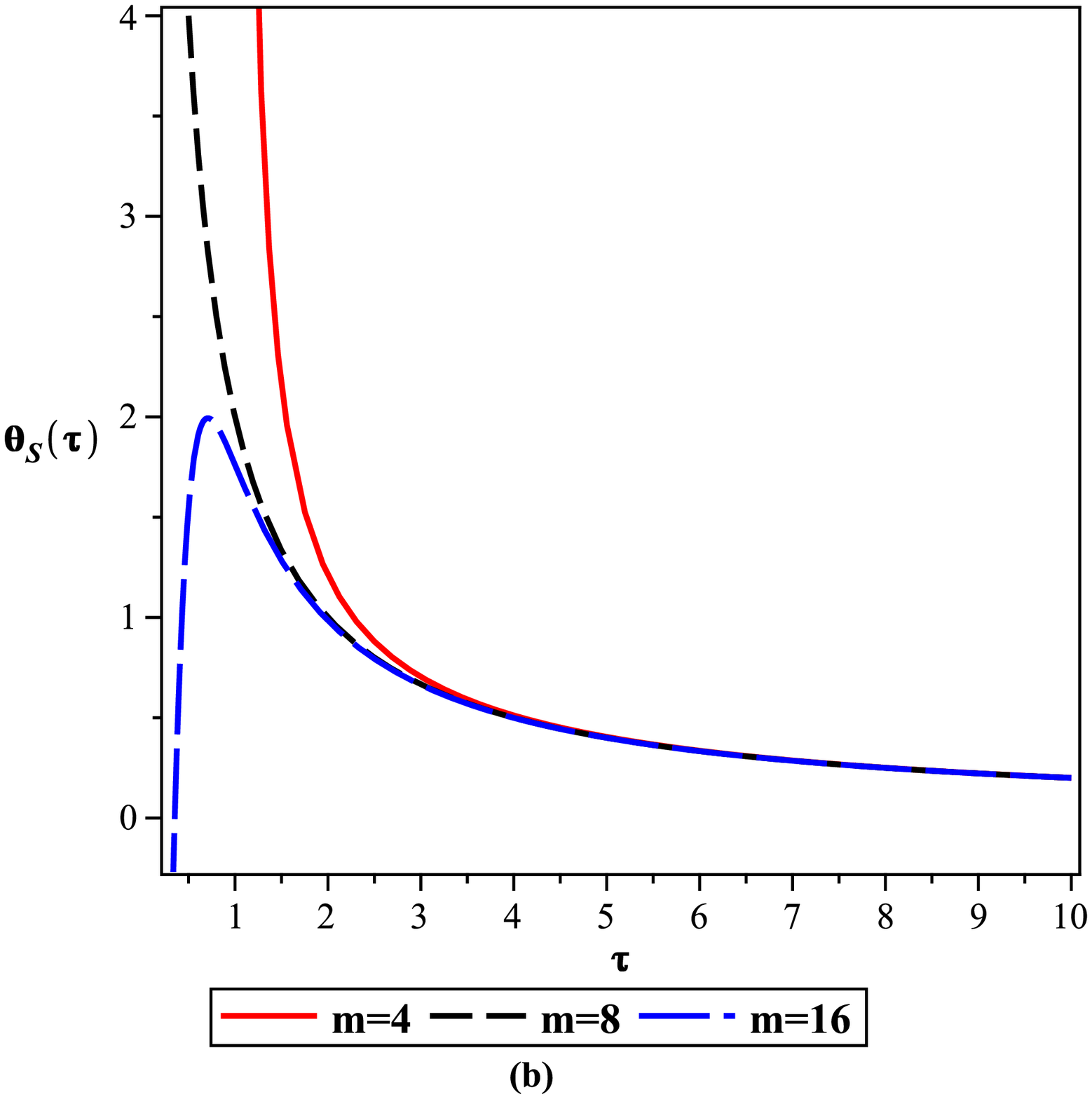}
\vspace{0.5cm}
\caption{Quantitative effect of variation of charge (in fig.(a)) and mass (in fig.(b)) for stringy BH; (a) $m=1$, $\lambda=1$, ( figure at right side of fig.(a) with no number represents the zoom out section of fig.(a) near $\tau=0$ ); (b) $q_{_S}=1$, $\lambda=1$.}
\protect\label{f2}
\end{figure}

\begin{figure}[h]
\includegraphics[width=3.8cm,height=4cm]{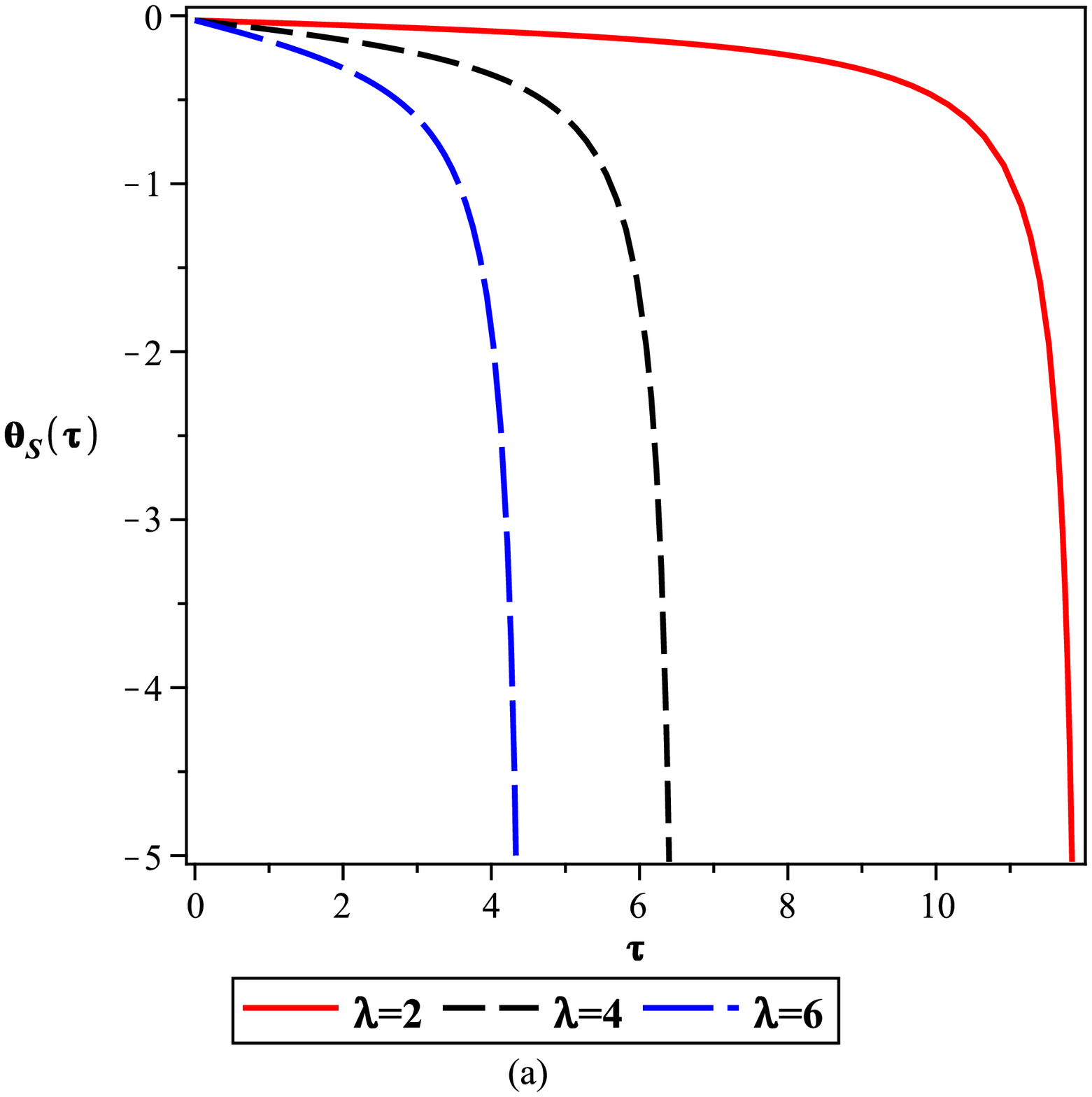}
\hskip1mm \includegraphics[width=3.8cm,height=4cm]{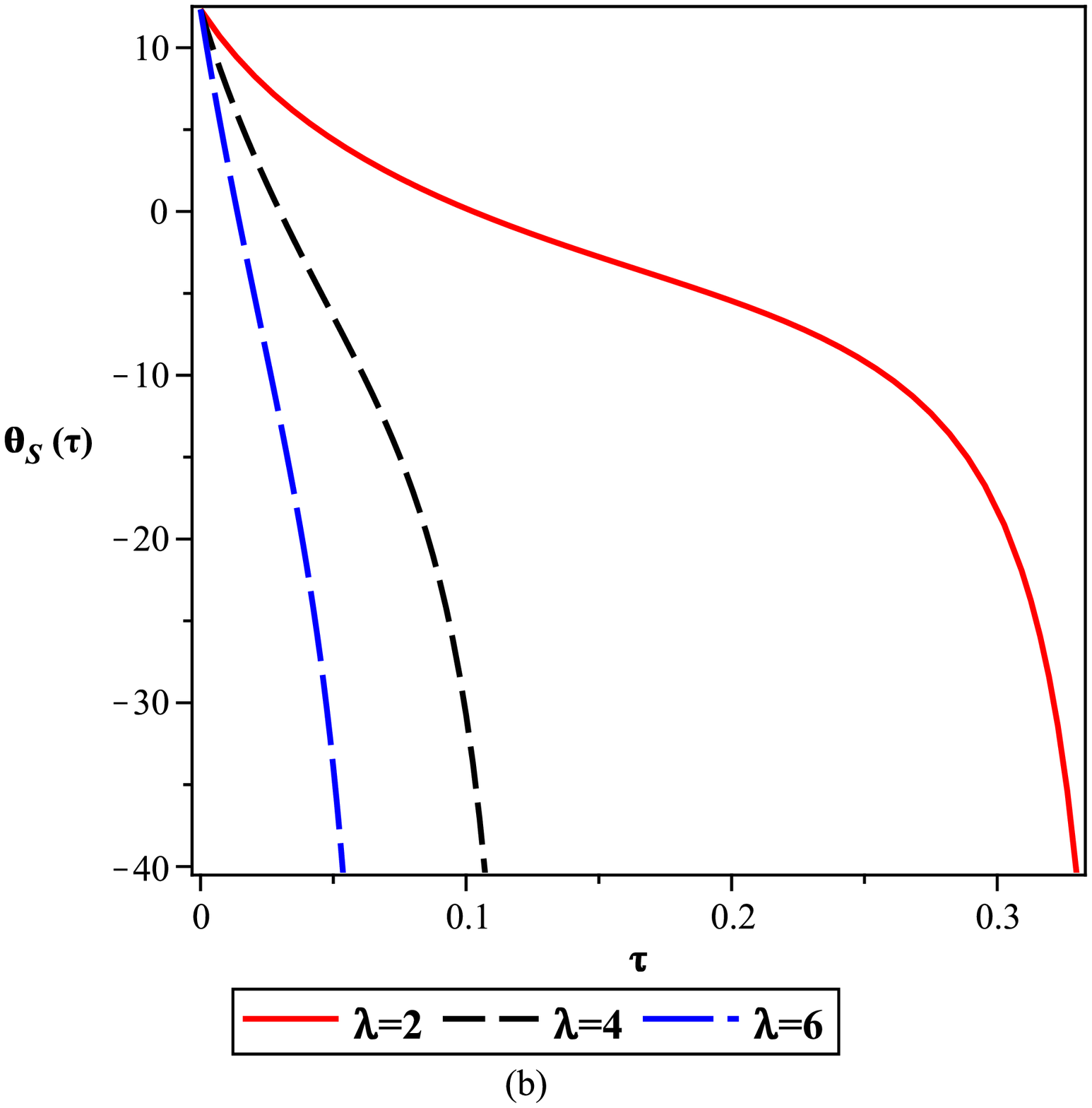}
\hskip1mm \includegraphics[width=3.8cm,height=4cm]{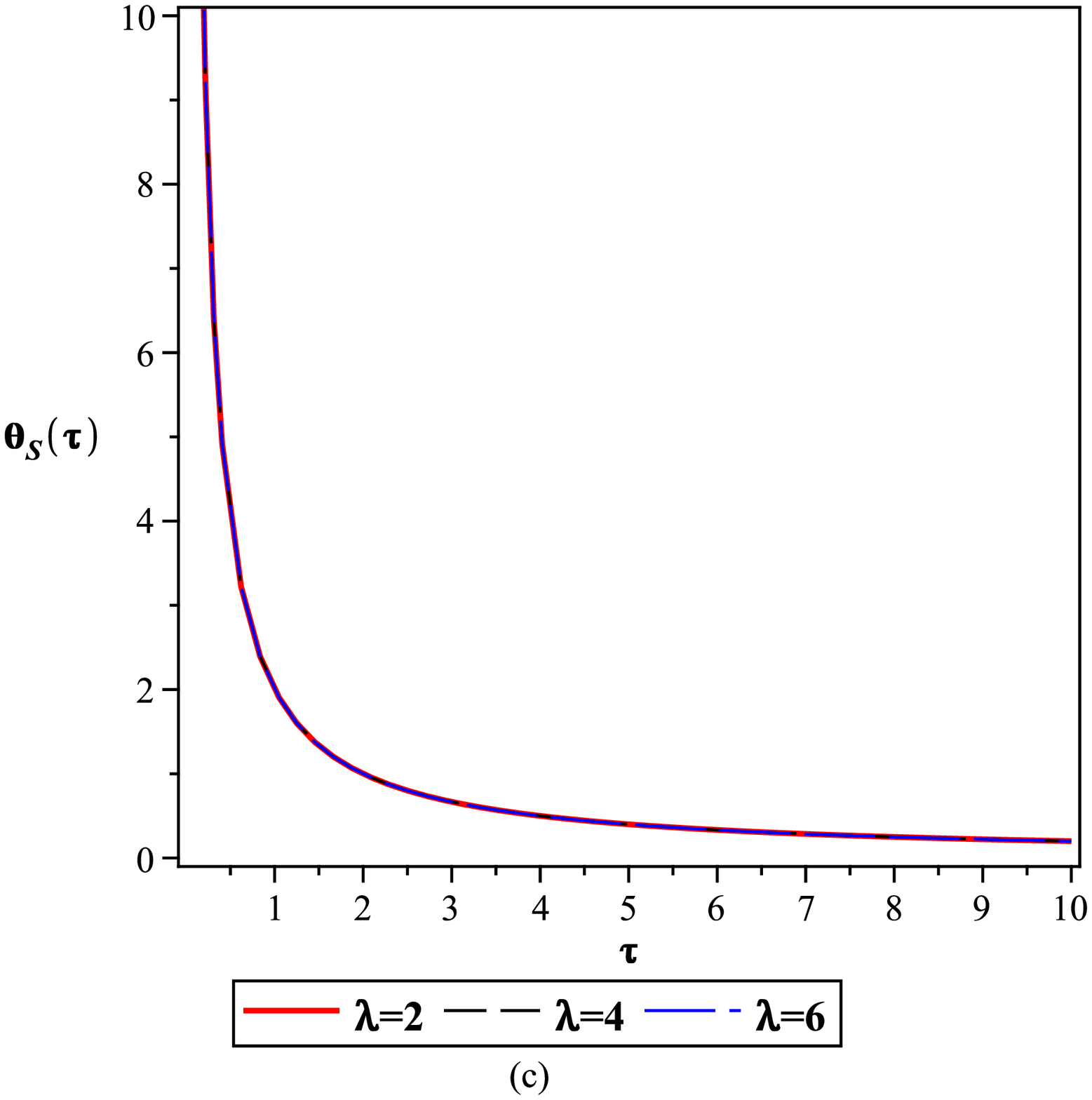}
\vspace{0.5cm}
\caption{Quantitative effect of variation of cosmological constant for stringy BH with,(a) $q_{_S}=100$, $m=1$, $\lambda =1$; (b) $q_{_S}=m=1$, $\lambda =1$; (c) $q_{_S}=1$, $m=100$, $\lambda=1$.}
\protect\label{f3}
\end{figure}

\begin{figure}[h]
\includegraphics[width=3.8cm,height=4cm]{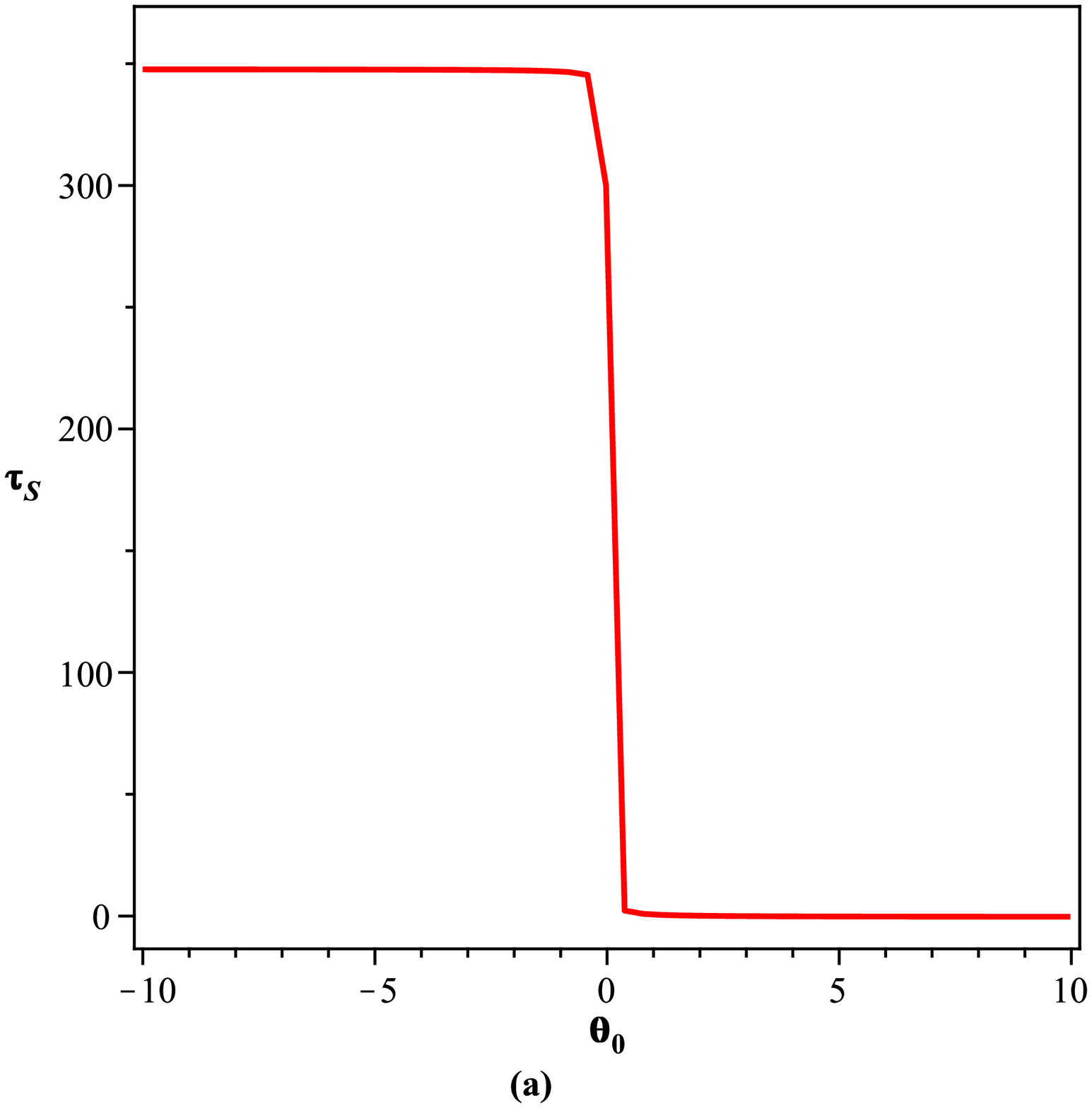}
\hskip1mm \includegraphics[width=3.8cm,height=4cm]{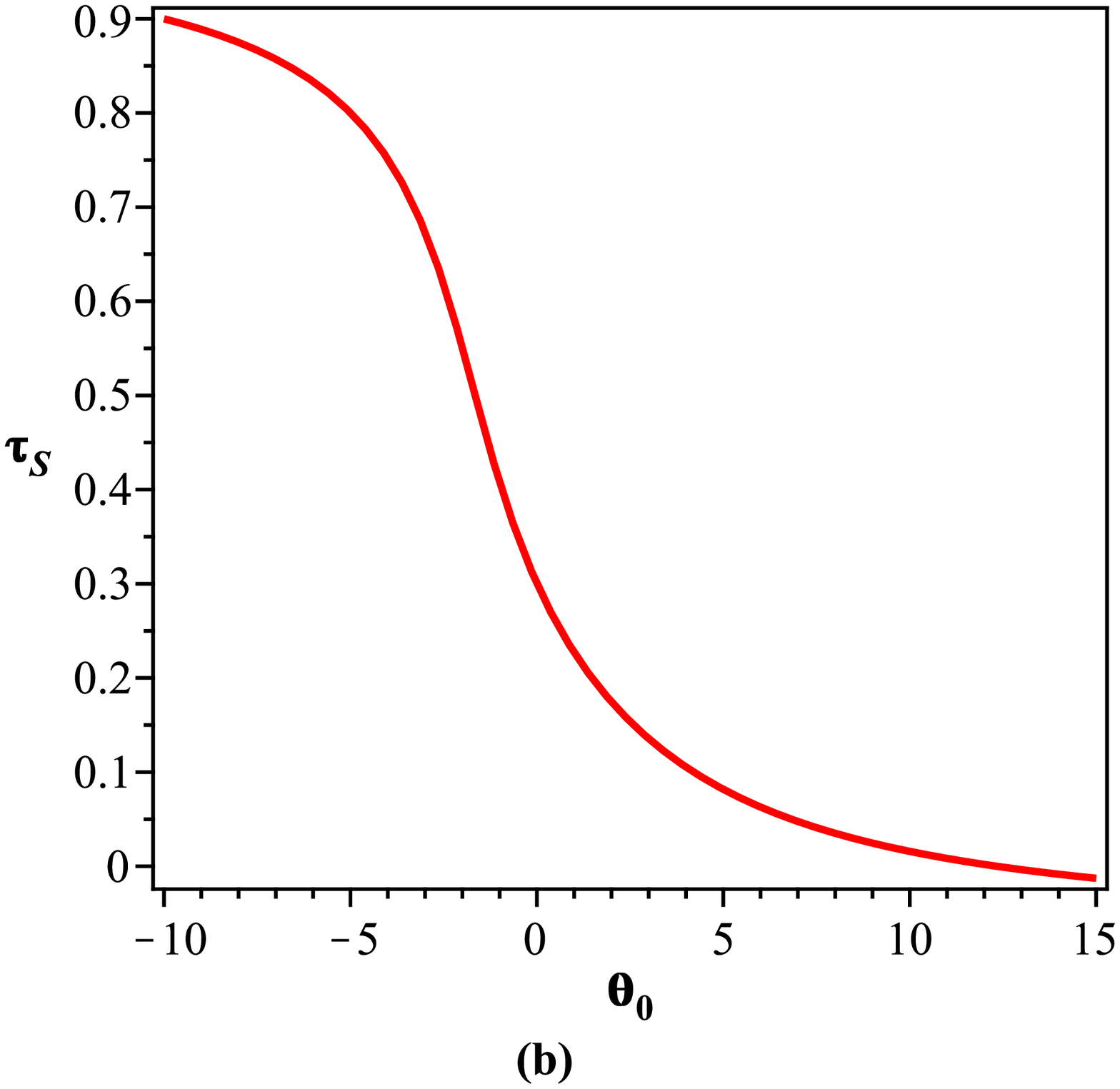}
\hskip1mm \includegraphics[width=3.8cm,height=4cm]{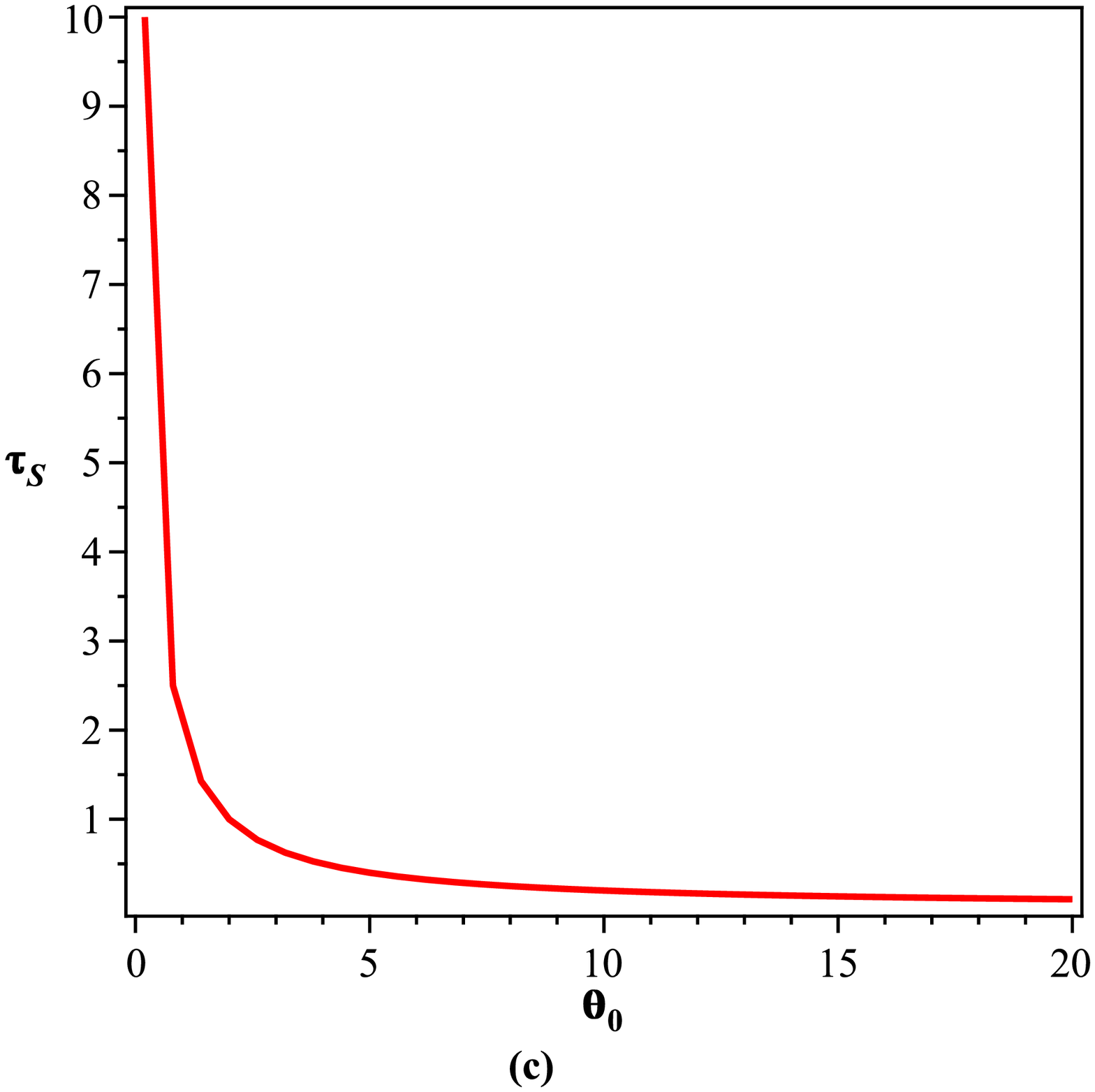}
\vspace{0.5cm}
\caption{Variation of time to singularity ($\tau_{_S}$) with initial expansion scalar ($\theta_0$) for stringy BH with, (a) $q_{_S}=100$, $m=1$, $\lambda =1$; (b) $q_{_S}=m=1$, $\lambda =1$; (c) $q_{_S}=1$, $m=100$, $\lambda=1$.}
\protect\label{f4}
\end{figure}

The occurrence of finite time singularity can be obtained from the exact solution of expansion scalar by solving the corresponding Raychaudhuri equation.
From eq.(\ref{eqn:sol_for stringy_qqqm}) (i.e. the solution of Raychaudhuri equation for $q_{_S}>>m$), the critical value of expansion scalar can be expressed in terms of initial values of various BH parameters as,
\begin{equation}
\theta^{c}_{0}= \frac{16\hspace{0.1cm}C_1\hspace{0.1cm}\lambda^4\hspace{0.1cm}m^4\hspace{0.1cm}\tau_0^4 + 24\hspace{0.1cm}K\hspace{0.1cm}m^2\hspace{0.1cm}q^2_{_S}\hspace{0.1cm}\lambda^2\hspace{0.1cm}\tau_0 - 3\hspace{0.1cm}C_1\hspace{0.1cm}q^4_{_S}}{L\hspace{0.1cm}N}
\label{eqn:critical_theta for stringy}
\end{equation}
where,
\begin{eqnarray}
 K = (C_1\hspace{0.1cm}\tau_0 - \frac{4}{3}),\hspace{0.5cm} L = 4\lambda^2\hspace{0.1cm}m^2\hspace{0.1cm}\tau_0^2 + q^2_{_S},
\nonumber
\end{eqnarray}
\begin{equation}
N = 4\hspace{0.1cm}C_1\hspace{0.1cm}m^2\hspace{0.1cm}\lambda^2\hspace{0.1cm}\tau_0^3 - 12\hspace{0.1cm}\lambda^2\hspace{0.1cm}m^2\hspace{0.1cm}\tau_0^2 - 3\hspace{0.1cm}C_1\hspace{0.1cm}q^2_{_S}\hspace{0.1cm}\tau_0 + q^2_{_S}.
\end{equation}
It can be observed from eq.(\ref{eqn:critical_theta for stringy}) that $\theta\rightarrow\infty$ as $N\rightarrow0$, hence $C_1$ can be fixed from this condition as,
\begin{equation}
{C_1}=\frac{12\hspace{0.1cm}{\tau^2_0\hspace{0.1cm}\lambda^2\hspace{0.1cm}m^2 - q^2_{_S}}}{4\hspace{0.1cm}\tau^3_0\hspace{0.1cm}\lambda^2\hspace{0.1cm}m^2 - 3\hspace{0.1cm}\tau_0\hspace{0.1cm}q^2_{_S}}.
\label{eqn:C_stringy_qqqm}
\end{equation}
For a particular set of parameters used in eq.(\ref{eqn:C_stringy_qqqm}) (with $q_{_S}=100$, $m=1$, $\lambda=1$, $\tau_0=20$), $C_1$ is calculated and used in eq.(\ref{eqn:critical_theta for stringy}) to obtain the critical value of expansion scalar as, $\theta_{0}^c(300)=-0.02122$.
For $\theta_{0}<\theta^{c}_{0}$, geodesic focusing occurs (see fig.(\ref{f1})).
A similar analysis for $m=q_{_S}$ (with $q_{_S}=1$, $m=1$, $\lambda=1$), results in $\theta_{0}^c(0.5)=-1.65517$, while for $m>>q_{_S}$ (with $q_{_S}=1$, $m=100$, $\lambda=1$) $\theta_{0}^c(1)=2$.
For $m>>q_{_S}$ geodesic defocusing is observed if $\theta_{0}>\theta^{c}_{0}$, while geodesic focusing does not occur otherwise.
Hence it is clear that evolution of geodesics depends on initial conditions of expansion scalar critically.
\\
\noindent In fig.(\ref{f1}) geodesic focusing is observed for $q_{_S}>>m$, $q_{_{RN}}>>m$ and $q_{_S}=m$, $q_{_{RN}}=m$ cases, while this effect is absent if  $m>>q_{_S}$, $m>>q_{_{RN}}$.
Fig.(\ref{f1}(f)) shows that in the absence of charge $q_{_{RN}}$, dotted curve represents the behaviour of an analogue to spherically reduced $2$D Schwarzschild BH spacetime.
Due to stringy corrections made in the action, corresponding charged stringy BH spacetime does not reduced to Schwarzschild BH spacetime in the absence of the charge $q_{_S}$.
Fig.(\ref{f2}) represents quantitative effect of variation of BH parameters such as charge and mass of stringy BH on the evolution of expansion scalar.
Variation of parameters does not effect the evolution of expansion scalar qualitatively, but only the critical value of expansion scalar for focusing(/defocusing) is changed correspondingly.\\
Fig.(\ref{f3}) represents the effect of variation of the cosmological constant $\lambda$ on the evolution of expansion scalar, it is observed for $q_{_S}>>m$ and $q_{_S}=m$ that geodesics focus at smaller values of affine parameter as $\lambda$ is increased. Variation of $\lambda$ does not effect the evolution if $m>>q_{_S}$.\\
The variation of time to singularity(\citeauthor{das09} \citeyear{das09}) ($\tau_{_S}$) versus initial value of expansion scalar ($\theta_0$) for stringy BH is presented in Fig.(\ref{f4}).
Time to singularity reduces as mass dominates over charge, again this figure also justifies the absence of focusing for $m>>{q_{_S}}$.
\begin{figure}[h]
\includegraphics[width=3.8cm,height=4cm]{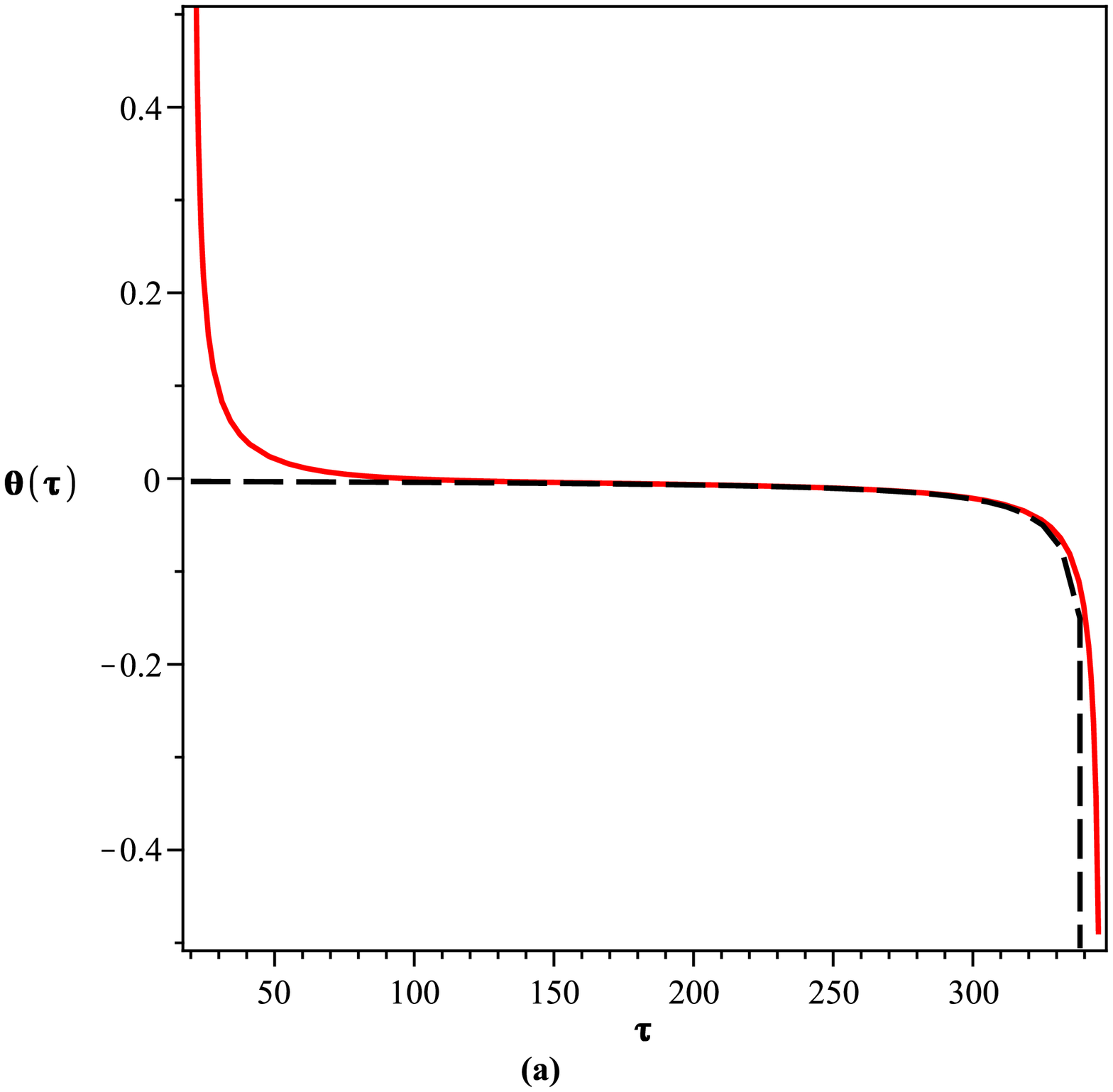}
\hskip2mm\includegraphics[width=3.8cm,height=4cm]{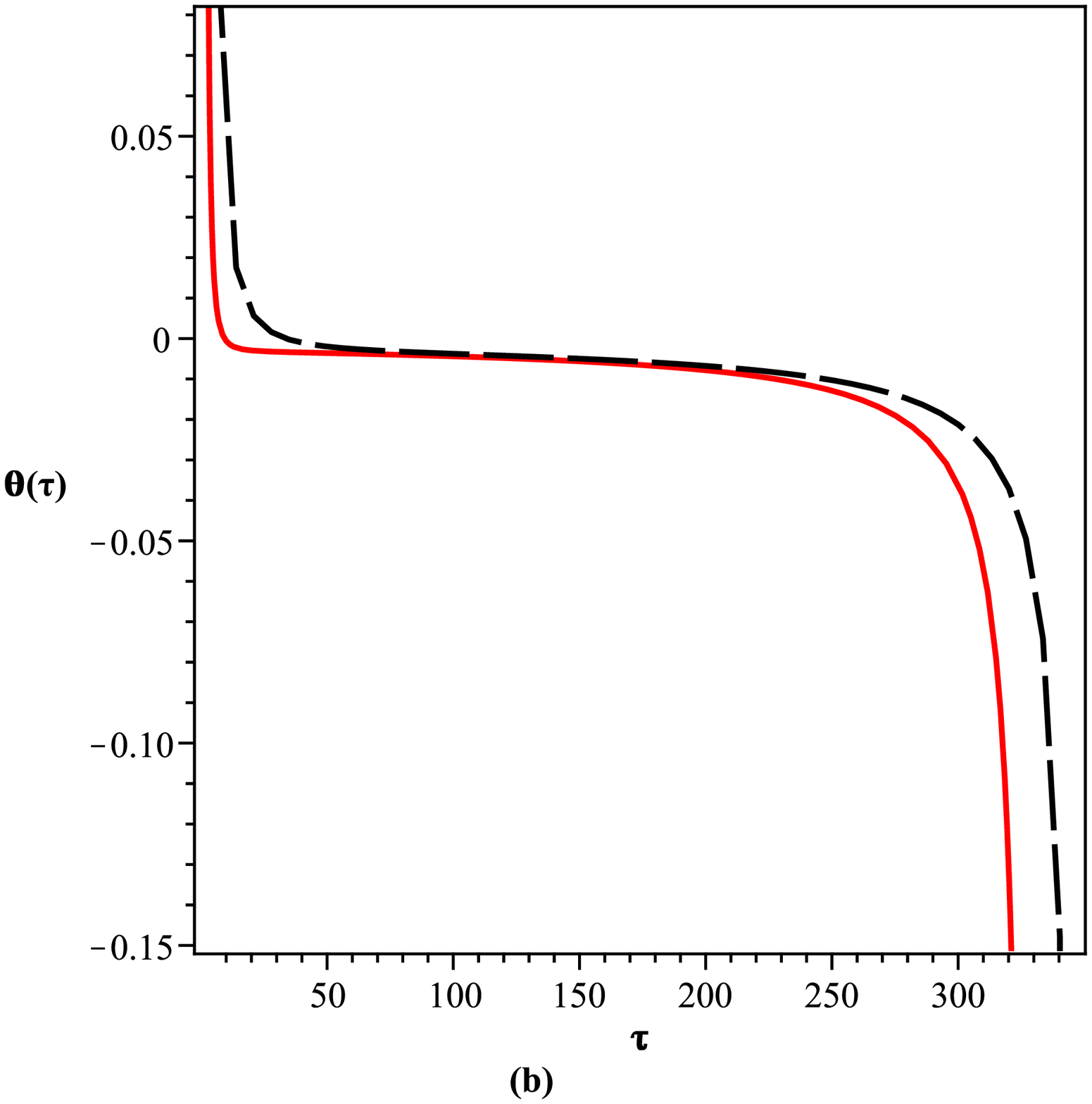}
\hskip2mm\includegraphics[width=3.8cm,height=4cm]{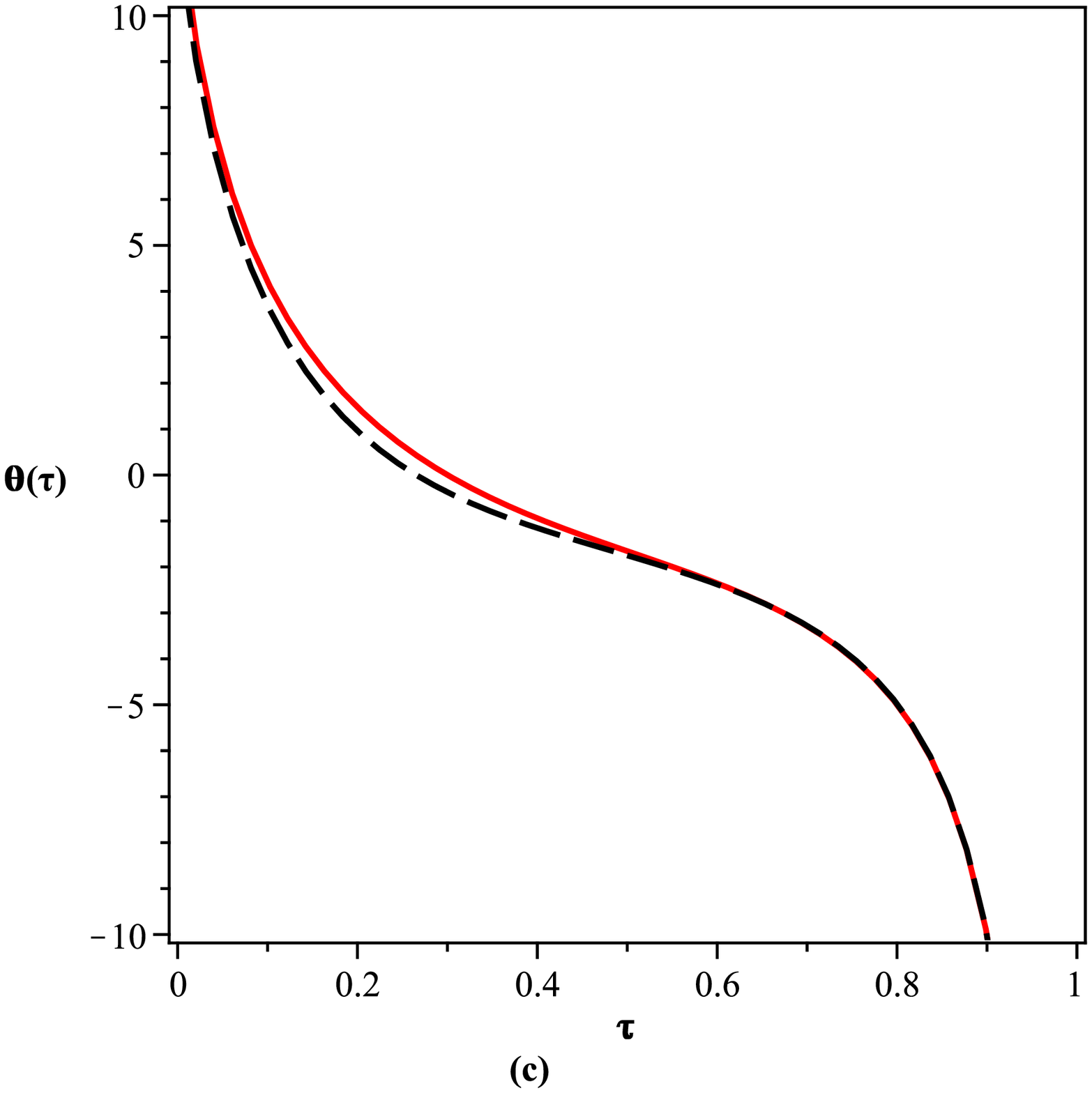}
\hskip2mm\includegraphics[width=3.8cm,height=4cm]{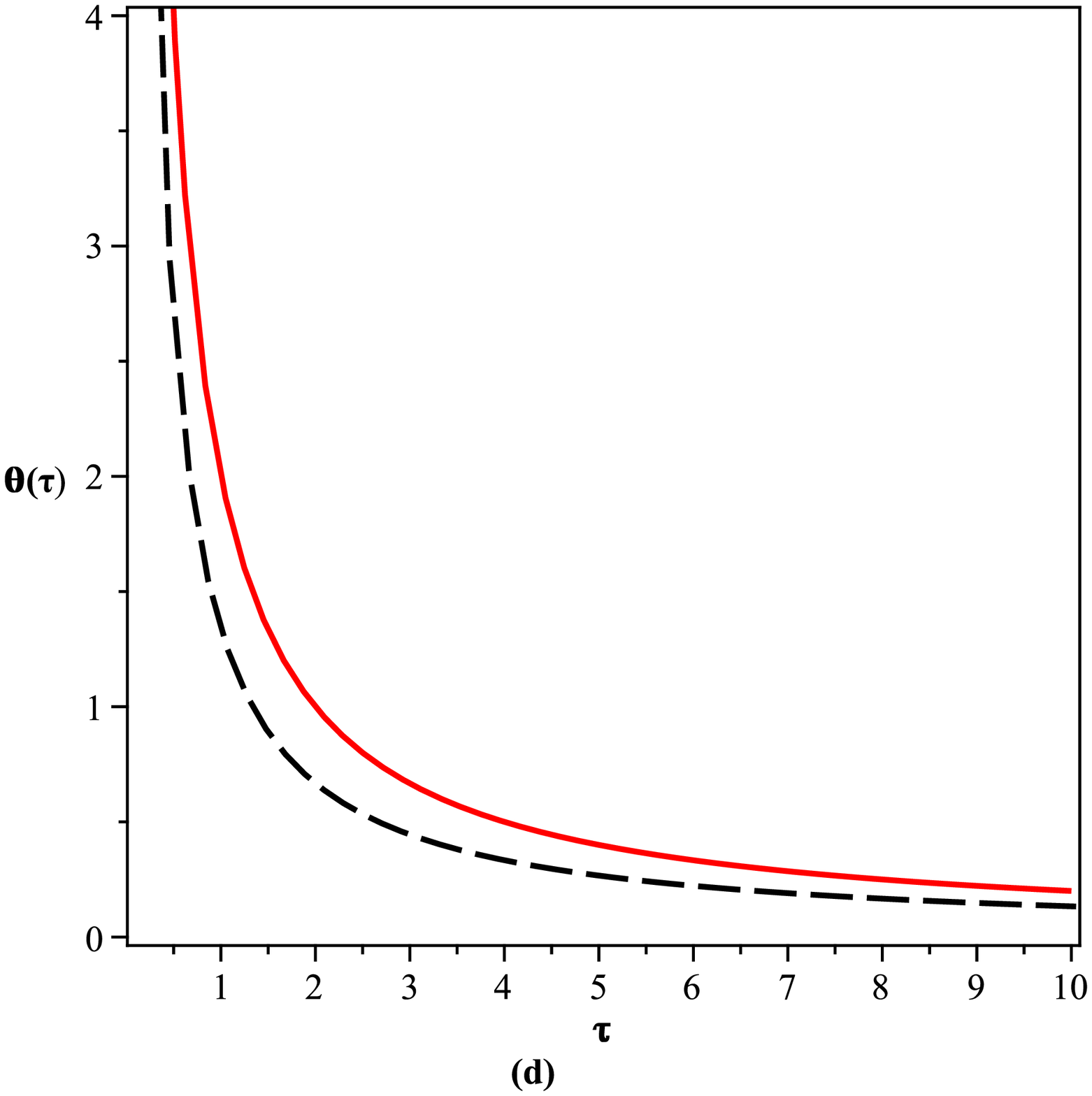}
\nonumber
\end{figure}
\newpage
\newpage
\newpage
\section{Discussion and Conclusion}
In the present work, we have studied the evolution of timelike geodesics in the background of charged $2$D stringy and Reissener-Nordstr$\ddot{o}$m BH spacetimes.
The conclusions are summarized as below:
\vspace{-2.5mm}
\begin{itemize}
\item The evolution of expansion scalar for a charged stringy BH and Reissener-Nordstr$\ddot{o}$m BH arising in GR in $2$D is discussed in detail.
The expression of expansion scalar and the conditions for occurrence of geodesic focusing/defocusing are examined using corresponding Raychaudhuri equations for expansion scalar with different settings of BH parameters.
The effect of charge and cosmological constant on the evolution of geodesic congruences is also analysed.
\item Geodesic focusing is observed in the cases of naked singularity (i.e. when charge dominates over mass) and extremal BHs (i.e. when charge and mass are equal), while it is observed that focusing is absent when mass dominates over charge for both the cases.
The results obtained for stringy BH case are in agreement with the results of our previous work (\citeauthor{uni14} \citeyear{uni14}) in view of the solutions of geodesic deviation equation.
\item By solving Raychaudhuri equation for expansion scalar as an \textit{initial value problem}, it is clear that initial values of expansion scalar plays a critical role in geodesic focusing/defocusing (wherever observed) in each case.
It is observed from the evolution of expansion scalar that initially converging geodesics converge more rapidly in comparison to initially diverging ones.
The critical values of expansion scalar for stringy BH case are also calculated for $q_{_S}>>m$, $q_{_S}=m$ and $q_{_S}<<m$ using different values for concerned parameters.
It is observed that for $q_{_S}>>m$ and $q_{_S}=m$ the geodesic focusing occurs if $\theta_{0}<\theta^{c}_{0}$, while for $m>>q_{_S}$ geodesic defocusing is observed if $\theta_{0}>\theta^{c}_{0}$.
\item The pattern of geodesic focusing is found qualitatively similar for the cases $q_{_S}>>m$ and $q_{_S}=m$ for different values of the cosmological constant while geodesic focusing is not observed in case of $m>>q_{_S}$.
However the scale of geodesic focusing for the abovementioned cases differs quantitatively.
Further a slightly positive increment in cosmological constant value assists the scale of geodesic focusing such that focusing occurs  more rapidly.
\item One can observe for stringy BH background that time to singularity ($\tau_{_S}$) decreases considerably for extremal case (i.e. $q_{_S}=m$) in comparison to naked singularity (i.e. $q_{_S}>>m$).
It is also clear from the variation of $\tau_{_S}$ with initial value of the expansion scalar ($\theta_{0}$) that geodesic focusing is absent for $m>>q_{_S}$ case.
The comparative study of $2$D charged stringy BH and Reissener-Nordstr$\ddot{o}$m BH shows that the qualitative behaviour of the evolution of the expansion scalar is similar in both the cases while it differs quantitatively.
\end{itemize}
In order to have more accurate visualization of the kinematics of geodesic flows in the background of different charged BH spacetimes, we intend to investigate the Reissener-Nordstr$\ddot{o}$m BH spacetime in $4$D and its charged analogue in other alternative theories of gravity in future.
The investigations made in this work can play an important role in addressing the issues related with the formation of accretion disk around BHs and is a matter of separate discussion.
It would also be worth studying the nature of null congruences in a similar fashion along with the notions of gravitational lensing.
\section*{Acknowledgements}
The authors RSK and HN would like to thank Department of Science and Technology, New Delhi for financial support through grant no. SR/FTP/PS-31/2009. The authors are thankful to the anonymous referee for his valuable suggestions. \\
\vspace{1mm}
\textbf{Conflict of Interest: The authors declare that they have no conflict of interest.}

\newpage
\section*{Appendix I}
Non-zero components of the Christoffel symbols and Ricci scalar:\\
\flushleft{\textbf{Case I}:}
\begin{eqnarray}
\Gamma^0_{01}=-\frac{2\lambda m e^{-2\lambda r}-q_{_S}^2e^{-2\lambda r}}{-1+2me^{-2\lambda r}-q_{_S}^2e^{-4\lambda r}}
\nonumber
\\
\Gamma^1_{00}=-2\lambda(-1+2me^{-2\lambda r}-q_{_S}^2e^{-4\lambda r})(me^{-2\lambda r}-q_{_S}^2e^{-4\lambda r})
\nonumber
\\
\Gamma^1_{11}=\frac{2\lambda(me^{-2\lambda r}-q_{_S}^2e^{-4\lambda r})}{-1+2me^{-2\lambda r}-q_{_S}^2e^{-4\lambda r}}
\nonumber
\\
R=8\lambda^2(me^{-2\lambda r}-2q^2_{_S}e^{-4\lambda r})
\nonumber
\end{eqnarray}
\flushleft{\textbf{Case II}:}
\begin{eqnarray}
\Gamma^{0}_{01}=-{\frac {mr-{q^{2}_{_{RN}}}}{r \left( -{r}^{2}+2\,mr
-q^{2}_{_{RN}} \right) }}
\nonumber
\\
\Gamma^{1}_{00}=-\frac{1}{r^5}(-r^2+2mr-q^2_{_{_{RN}}})(mr-q^2_{_{_{RN}}})
\nonumber
\\
\Gamma^{1}_{11}={\frac {mr-q^{2}_{_{RN}}}{r \left( -{r}^{2}+2\,mr-{
q^{2}_{_{RN}}} \right) }}
\nonumber
\\
R=\frac{2}{r^4}(2mr-3q^2_{_{RN}})
\nonumber
\end{eqnarray}

\bibliography{biblio-u1}

\begin{thebibliography}{99}
\bibitem[Garfinkle et al. (1991)]{gar91} Garfinkle, D., Horowitz, G.T., and Strominger, A.: \textit{Charged black holes in string theory}. \textit{Phys. Rev.} \textbf{D43}, 3140 (1991).
\bibitem[Campbell et al. (1990)]{cam90} Campbell, B. A., Duncan, M. J., Kaloper, N., and Olive, K. A.: \textit{Axion hair for Kerr black holes }. \textit{Phys. Lett.} \textbf{B251}, 34 (1990).
\bibitem[Shapere et al. (1991)]{shp91} Shapere, A. , Trivedi, S. , Wilczek, F. : \textit{Dual dilaton dyons}. \textit{Mod. Phys. Lett.} \textbf{A6}, 2677 (1991).
\bibitem[Bekenstein (1972)]{bek72} Bekenstein,J. D.: \textit{Nonexistence of baryon number for static black holes}. \textit{Phys. Rev.} \textbf{D5}, 1239 (1972).
\bibitem[Chase (1970)]{cha70} Chase,J. E. : \textit{Event horizons in static scalar-vacuum space-times}. \textit{Comm. Math. Phys.} \textbf{19}, 276 (1970).
\bibitem[Mignemi and Stewart (1993)]{mig93} Mignemi, S. and  Stewart, N. R. : \textit{Charged black holes in effective string theory}. \textit{Phys. Rev.} \textbf{D47}, 5259 (1993).
\bibitem[Poisson (2004)]{poi04} Poisson, E. : \textit{{A relativists' toolkit: The mathematics of Black hole mechanics}}. (Cambridge University press, Cambridge, 2004).
\bibitem[Wald (1984)]{wal84} Wald, R. M. :\textit{General Relativity}. (University of Chicago Press, Chicago, USA, 1984).
\bibitem[Hartle (2003)]{har03} Hartle, J. B. : \textit{Gravity: An Introduction To Einstein's General Relativity}. (Pearson Education Inc., Singapore, 2003).
\bibitem[Schutz (1985)]{sch85} Schutz, B. F. : \textit{A First Course in General Relativity'}. (Cambridge Uni. Press, Cambridge, 1985).
\bibitem[Chandrasekhar (1983)]{cha83} Chandrasekhar, S. : \textit{The Mathematical Theory of Black Holes}. (Oxford Uni. Press, New York, 1983).
\bibitem[Joshi (1997)]{jos97} Joshi, P. S. : \textit{Global aspects in gravitation and cosmology}. (Oxford University Press, Oxford, UK, 1997).
\bibitem[Vagenas (2003)]{vag03} Vagenas, E. C. : \textit{Energy Distribution in $2$d Stringy Black Hole Backgrounds}. \textit{JHEP} 0307 046 (2003) (UB-ECM-PF-03/07).
\bibitem[E.C. Vagenas (2003)]{vag3} Vagenas, E. C. : \textit{$2$d Stringy Black Holes and Varying Constants}. \textit{Int. J. Mod. Phys. A }\textbf{18} 5781 (2003) (UB-ECM-PF-03/16).
\bibitem[Dasgupta et al. (2009)]{das09} Dasgupta, A. , Nandan, H., Kar, S. : \textit{Kinematics of geodesic flows in stringy Black Hole backgrounds}. Phys. Rev. \textbf{D79}, 124004 (2009) (and references therein).
\bibitem[Dasgupta et al. (2008)]{das08} Dasgupta, A. , Nandan, H. , and Kar, S. : \textit{Kinematics of deformable media}. Annals. Phys. \textbf{323}, 1621 (2008).
\bibitem[Dasgupta et al. (2009)]{das9} Dasgupta, A. , Nandan, H. , and Kar, S. : \textit{Kinematics of flows on curved, deformable media}. Int. J. Geom. Meth. Mod. Phys. \textbf{6}, 645 (2009).
\bibitem[Ghosh et al. (2010)]{gho10} Ghosh, S. , Kar, S. , and Nandan, H. : \textit{Confinement of test particles in warped spacetimes}. Phys. Rev. \textbf{D85}, 024040 (2010).
\bibitem[Fernando (2012)]{fer12} Fernando, S. : \textit{Schwarzschild black hole surrounded by quintessence: Null geodesics}. \emph{ Gen. Rel. Grav.} \textbf{44}, 1857 (2012).
\bibitem[Pugliese (2011)]{pug11} Pugliese, D. , Quevedo, H. , and Ruffini, R. : \textit{Motion of charged test particles in Reissner-Nordstr$\ddot{o}$m spacetime}. \textit{Phy. Rev.} \textbf{D83}, 104052 (2011).
\bibitem [Koley et al. (2003)]{kol03} Koley, R. , Pal, S. , and Kar, S. : \textit{Geodesics and geodesic deviation in a two-dimensional black hole}. \textit{Am. J. Phys.} \textbf{71}, 1037 (2003).
\bibitem [Uniyal et al. (2014)]{uni14} Uniyal, R. , Nandan, H. , and Purohit, K. D. : \textit{Geodesic Motion in a Charged $2$D Stringy Blackhole Spacetime}. \textit{Mod.Phys.Lett.} \textbf{A29}, 1450157 (2014).
\bibitem [R.Uniyal et al. (2014)]{runi14} Uniyal, R. , Chandrachani Devi, N. , Nandan, H. , and Purohit, K. D. : \textit{Geodesic Motion in Schwarzschild Spacetime Surrounded by Quintessence}. \textbf{arXiv: 1406.3931 [gr-qc] } (2014).
\bibitem[Kar (2008)]{kar08} Kar, S. : \textit{An Introduction to the Raychaudhuri Equations}. \textit{Resonance J.Sci.Educ.} \textbf{13}, 319 (2008).
\bibitem[Kar and sengupta (2007)]{kar07} Kar, S. , and Sengupta, S. : \textit{The Raychaudhuri Equations:A brief review}. \textit{Pramana} \textbf{69}, 49 (2007).
\bibitem[Dadhich (2005)]{dad05} Dadhich, N. : \textit{Derivation of the Raychaudhuri Equation}. \textbf{arXiv:gr-qc/0511123v2} (2005).
\bibitem[Ehlers (2007)]{ehl07} Ehlers, J. : \textit{A. K. Raychaudhuri and his equation}. \textit{Pramana} \textbf{69}, 7 (2007).
\bibitem[McGuigan et al. (1992)]{mcg92} McGuigan, M. D. , Nappi, C. R. , and Yost, S. A. : \textit{Charged Black holes in Two-Dimensional String Theory}. \textit{Nucl. Phys.} \textbf{B375}, 421 (1992).
\bibitem[Sen (1992)]{sen92} Sen, A. : \textit{Rotating Charged Black hole Solution in Heterotic String Theory}. \textit{Phys. Rev. Lett.} \textbf{69}, 1006 (1992).
\bibitem[Kar (1997)]{kar97} Kar, S. : \textit{Stringy Black holes and Energy Conditions}. \textit{Phys.Rev.} \textbf{D55}, 4872 (1997).
\bibitem[Dasgupta et al. (2012)]{das12} Dasgupta, A. , Nandan, H. and Kar, S. : \textit{Geodesic flows in rotating Black hole backgrounds}. \textit{Phys. Rev.} \textbf{D85}, 104037 (2012).
\bibitem[Weinberg (2004)]{wen04} Weinberg, S. : \textit{Gravitation and Cosmology: Principles and Applications of General Theory of Relativity}. (Jhon Wiley and Sons(Aisa), Singapore, 2004).
\bibitem[Ovidiu-Cristinel (2012)]{ovi12} Ovidiu-Cristinel Stoica, : \textit{Analytic Reissner-Nordstr$\ddot{o}$m Singularity}. \textit{Phys.Scripta} \textbf{85}, 055004 (2012).
\bibitem[Townsend (1997)]{tow97} Townsend, P. K. : \textit{Black holes: Lecture notes}. \textbf{arXiv: gr-qc/9707012v1} (1997).
\bibitem[Mandal et al. (1991)]{man91} Mandal, G. , Sengupta, A. M. , and Wadia, S. R. : \textit{Classical solution of 2-dimensional string theory}. \textit{Mod, Phys. Letts. }\textbf{A6}, 1685 (1991).
\end{thebibliography}


\end{document}